\documentstyle[12pt]{article}
\makeatletter
\def\@maketitle{\newpage
 \null
 {\normalsize \tt \begin{flushright}
  \begin{tabular}[t]{l} \@date
  \end{tabular}
 \end{flushright}}
 \begin{center}
 \vskip 2em
 {\LARGE \@title \par} \vskip 1.5em {\large \lineskip .5em
 \vfill
 \begin{tabular}[t]{c}\@author
 \end{tabular}\par}
 \end{center}
 \par
 \vskip 1.5em
 \vfill}
\makeatother
\makeatletter
\@addtoreset{equation}{section}
\makeatother

\newcommand{\VEV}[1]{\langle #1 \rangle}

\newcommand{\Lag}{\mbox{${\cal L}$}}
\newcommand{\fsl}[1]{\not \kern-2pt #1}

\newcommand{\GeV}{\hbox{GeV}}
\newcommand{\MeV}{\hbox{MeV}}
\newcommand{\lesssim}{{<\atop\sim}}
\newcommand{\xbar}[1]{#1 \hspace{-5.5pt}/}

\newcommand{\SxSB}{\mbox{$S\chi SB$}}

\def\fsl#1{\setbox0=\hbox{$#1$}           
   \dimen0=\wd0                                 
   \setbox1=\hbox{/} \dimen1=\wd1               
   \ifdim\dimen0>\dimen1                        
      \rlap{\hbox to \dimen0{\hfil/\hfil}}      
      #1                                        
   \else                                        
      \rlap{\hbox to \dimen1{\hfil$#1$\hfil}}   
      /                                         
   \fi}                                         %

\title{  Tightly Bound Composite Higgs\thanks{
    Work supported in part by the Takeda Science Foundation and the
    Ishida Foundation, and by the International Collaboration Program
    of the Japan Society for the Promotion of Science.
  }%
\ \thanks{To appear in Proc. Workshop on Beyond the Standard Model III,
  Carleton University, Ottawa, Canada, June 22 - 24, 1992
  (World Scientific Pub. Co., Singapore, 1993).
 }
}
\author{
  Koichi Yamawaki\thanks{e-mail:b42060a@nucc.cc.nagoya-u.ac.jp
  }\\
  Department of Physics, Nagoya University \\
  Nagoya 464-01, Japan
}
\date{
  DPNU-92-49 \\
  November 1992
}
\begin{document}
\renewcommand{\thepage}{ }
\begin{titlepage}

\maketitle

\begin{abstract}
  \small
  \setlength{\baselineskip}{12pt}
  We explain general features of the tightly bound composite Higgs
  models proposed in recent years; walking technicolor, strong ETC
  technicolor,
  and a top quark condensate, etc.. These models are all characterized
  by the large anomalous dimension due to nontrivial short distance
  dynamics
  in the gauged Nambu-Jona-Lasinio models (gauge theories plus
  four-fermion interactions).
  \par
 \vspace{1.6cm}
\end{abstract}
\end{titlepage}
\renewcommand{\thepage}{\arabic{page}}
\section{Introduction}

It is well known that the Higgs sector in the standard model is
precisely
the same as the sigma model except that $\VEV\sigma=f_\pi=93\MeV$ is
simply replaced by the Higgs vacuum expectation value (VEV)
$=F_\pi=250\GeV$, roughly 2600 times
scale-up. Since the sigma model is a low energy effective theory of
QCD reproducing the spontaneous chiral symmetry breaking (\SxSB) due
to the
quark-antiquark pair condensate, one is naturally led to
speculate that there might exist a microscopic theory
for the Higgs sector, with the
Higgs VEV being replaced by the fermion-antifermion pair condensate
due to
yet another strong interaction called Technicolor (TC)\cite{kn:FS81}.

Unfortunately, the original version of TC was too naive to survive
the FCNC (flavor-changing neutral current) syndrome\cite{kn:FS81}.
It was not the end of the story, however. QCD-like theories
(simple scale-up's of QCD) turned out not to be the unique candidate
for the underlying dynamics of the Higgs sector. Actually, there have
been proposed varieties of composite Higgs models which, though
equally behaving
as the sigma model in the low energy region, still have different
high energy behaviors than QCD; Walking
Technicolor\cite{kn:Hold85,kn:YBM86},
Strong ETC Technicolor\cite{kn:MY89} and the Top Quark Condensate
(Top Mode Standard Model)\cite{kn:MTY89}, etc.. In this talk I would
like to give a general description of such new possibilities in terms
of the large
anomalous dimension, based on the explicit dynamical model, the gauged
Nambu-Jona-Lasinio (NJL) model (gauge theory plus four-fermion
interactions).

\section{Large Anomalous Dimension and Tightly Bound States}

Such a different high energy behavior than QCD simply reflects the
relatively short distance dynamics relevant to the composite Higgs.
Using the operator product expansion (OPE) and the renormalization-group
equation (RGE) for the condensed fermion propagator $iS^{-1}(p)
=\xbar{p} -\Sigma(-p^2)$ (in Landau gauge), we may write the dynamical
mass $\Sigma (-p^2)$ (``nonlocal order parameter'') in the high energy
region (we switch over to Euclidean momentum hereafter;
$-p^2 \rightarrow p^2$):
\begin{equation}
  \Sigma(p^2)\mathop{\simeq}^{p \gg F_\pi}
    {\VEV{\bar \psi \psi}_{\mu} \over -p^2} \cdot \exp \left[
     \int_0^{t} \gamma_m(t') dt' \right],
\label{eq:(1)}
\end{equation}
where $t \equiv \ln (p/\mu)$ and $\mu (=O(F_\pi))$ is the
renormalization point. The canonical scaling $1/p^2$ is modified due to
the anomalous dimension $\gamma_m (t)$. Similarly, the
fermion-antifermion
condensate (``local order parameter'') at a certatin high energy
scale $\Lambda$ takes
the form
\begin{equation}
 \VEV{\bar \psi \psi}_\Lambda = Z_m^{-1} \VEV{\bar \psi \psi}_{\mu}\\
 = \exp \left[ \int_0^{t_\Lambda} \gamma_m(t') dt' \right]
   \cdot \VEV{\bar \psi \psi}_{\mu} \simeq -\Lambda^2 \Sigma(\Lambda^2),
\label{eq:(2)}
\end{equation}
where $t_\Lambda \equiv \ln (\Lambda/\mu)$ and the last equation
follows from
comparison with Eq.(\ref{eq:(1)}) at $p^2 =\Lambda^2$.

Now, the ($N_f$-flavored) QCD has a logarithmically vanishing anomalous
dimension $\gamma_m \simeq A/2{\bar t}$, with $A = 24/(33-2N_f)$ and
$\bar t \equiv \ln(p/\Lambda_{QCD}) = t+\ln(\mu/\Lambda_{QCD})$,
which yields only a logarithmical correction (enhancement) to the
canonical one:
\begin{equation}
 \exp \left[ \int_0^t \gamma_m(t') dt' \right]
    \simeq \exp [\frac{A}{2} \ln (\bar t /{\bar t}_\mu)]
    = \left( \frac{\ln\frac{p}{\Lambda_{QCD}}}
                 {\ln\frac{\mu}{\Lambda_{QCD}}}
                 \right)^{A \over 2}.
\label{eq:(3)}
\end{equation}

On the other hand, if the theory has a non-vanishing anomalous dimension
$\gamma_m (t) \simeq \gamma_m \ne 0$ due to non-vanishing coupling
constant (behaving as a nontrivial ultraviolet (UV) fixed point/peudo
fixed point) at high energies, then we have a {\em power
enhancement} instead of the above logarithmic one:
\begin{equation}
 \exp \left[ \int_0^t \gamma_m(t') dt' \right]
    \simeq e^{\gamma_m t}
    = \left( \frac{p}{\mu}
      \right)^{\gamma_m}.
\label{eq:(4)}
\end{equation}
Accordingly, we have power-enhanced order parameters:
\begin{eqnarray}
 \Sigma(p^2) &\simeq& \frac{\VEV{\bar \psi \psi}_\mu}{-\mu^2}
                      \left( \frac{p}{\mu} \right)^{-2+\gamma_m},
                            \nonumber \\
 \VEV{\bar \psi \psi}_\Lambda
  &\simeq&  \left( \frac{\Lambda}{\mu}
              \right)^{\gamma_m}
            \cdot    \VEV{\bar \psi \psi}_{\mu}.
\label{eq:(5)}
\end{eqnarray}

This is actually the mechanism that Holdom\cite{kn:Hold81} proposed
{\em without explicit dynamics} to resolve the FCNC and the light pseudo
NG bosons problems of TC, by simply assuming $1 \lesssim \gamma_m$.
  In fact, in the extended technicolor
(ETC) scenario\cite{kn:FS81} with $\mu = \Lambda_{TC}$ and
$\Lambda = \Lambda_{ETC}$ ($\gg \Lambda_{TC}$), quarks/leptons mass is
given by
$
m \simeq \VEV{\bar \psi \psi}_{\Lambda_{ETC}}/{\Lambda_{ETC}}^2
$
and
$
m_{pseudo NG} \sim
\VEV{\bar \psi \psi}_{\Lambda_{ETC}}/\Lambda_{TC}\Lambda_{ETC}
$,
both of which are thus enhanced by the factor
$(\Lambda_{ETC}/\Lambda_{TC})^{\gamma_m}$.

Such a large enhancement also amplifies the small symmetry violation of
high
energy parameters. This fact was first utilized by Miransky, Tanabashi
and Yamawaki\cite{kn:MTY89,kn:Yama92} in the proposal of a top quark
condensate
($m_t \gg m_b$) and was re-emphasized in a slightly different
context\cite{kn:NNT91}.

A large anomalous dimension actually implies a tightly bound
Nambu-Goldstone (NG) bosons due to relatively short distance dynamics.
In fact, it is well known that the amputated Bethe-Salpeter amplitude
of the NG bosons at zero NG-boson-momentum is related to $\Sigma (p^2)$
through the Ward-Takahashi identity for
the axialvector vertex:
\begin{equation}
 \chi_\pi^a(p,p+q)|_{q^2=0} = \frac{1}{F_\pi}\tau^a\gamma_5 \Sigma(p^2)
 \sim \left( \frac{p}{\mu} \right)^{-2+\gamma_m}
\end{equation}
(in the $SU(2)_L \times SU(2)_R/SU(2)_V$ case). Thus in QCD with
$\gamma_m
\simeq 0$ we find $\chi_\pi \sim (p/\mu)^{-2}$ and hence the radius of
the
composite $\VEV{r} \simeq \mu^{-1} \simeq
F_\pi^{-1}$. In this case the sigma model description breaks down at the
order
of $O(F_\pi)$. On the other hand, in the extreme case of $\gamma_m
\simeq 2$
we have $\chi_\pi \sim {\rm const.}$ and hence $\VEV{r} \simeq
\Lambda^{-1}$
(almost point-like, or very tightly bound), in which case the sigma
model description persists up to the high energy scale $\Lambda$.

\section{Tightly Bound Composite Higgs Models}

\begin{flushleft}
  {\bf Walking Technicolor}\cite{kn:Hold85,kn:YBM86}
\end{flushleft}

Do such {\em explicit} dynamical models as have a large anomalous
dimension
really exist? The answer is ``yes''. It was first pointed out by
Yamawaki,
Bando and Matumoto\cite{kn:YBM86} that the technicolor within
the {\em ladder Schwinger-Dyson (SD) equation} (with the gauge coupling
constant {\em fixed} - non-runnnig) possesses an \SxSB\ {\em solution}
with a large anomalous dimension:
\begin{eqnarray}
 \gamma_m &\simeq& 1,\\
 \Sigma(p^2) &\sim& \frac{1}{p} \qquad (p \gg \Lambda_{TC}),\\
 \VEV{\bar \psi \psi}_{\Lambda_{ETC}} &\simeq&
  \left( \frac{\Lambda_{ETC}}{\Lambda_{TC}} \right)
 \cdot \VEV{\bar \psi \psi}_{\Lambda_{TC}},
\end{eqnarray}
and hence resolves the long standing problems of the old TC mentioned
above.
(Essentially the same observation was also made
by Akiba and Yanagida\cite{kn:YBM86} without notion of the anomalous
dimension.)

The above feature is actually the esssence of the ``walking TC'',
a generic name currently used (see Appelquist et al.\cite{kn:YBM86})
for a wider class of TC's with slowly
running ($A \gg 1$) gauge coupling {\em including} the non-running
($A \rightarrow \infty$, ``standing'') case
as an extreme case. In order for the walking TC be a realistic solution
of
the FCNC problem, however, it must be very close to the standing limit
anyway
(see Bando et al.\cite{kn:YBM86}). In the standing limit
the \SxSB\ solution exists only when the gauge coupling $\alpha
\equiv e^2/4\pi$ exceeds a critical value
$\alpha_c =\pi/(3C_F)$ ($C_F$: quadratic Casimir of the fermion
representation). Hence the critical coupling plays a role of a
nontrivial
UV fixed point. Thus the walking TC may be viewed as the TC with a
nontrivial UV fixed point/pseudo fixed point, with the coupling being
kept close to the critical coupling all the way up to $\Lambda_{ETC}$
scale.
It should be mentioned
that Holdom\cite{kn:Hold85} earlier recognized essentially the same
dynamics through a purely {\em numerical} analysis of the
ladder-type SD equation.

\begin{flushleft}
  {\bf Strong ETC Technicolor}\cite{kn:MY89}
\end{flushleft}

Next we come to the TC with even larger anomalous dimension,
$1 < \gamma_m <2$, which have been proposed\cite{kn:MY89},
based on the {\em explicit} dynamics of the gauged NJL model in the
framework
of the ladder SD equation\cite{kn:BLL86}, with the ETC/preonic
interactions
being simulated by the four-fermion interactions (see Yamawaki
et al.\cite{kn:YBM86}).
Actually, based on the \SxSB\ solution\cite{kn:KMY89},
we have a big enhancement of order
parameters due to a large anomalous dimension:
\begin{eqnarray}
 1 < \gamma_m &=& 1+\sqrt{1-\frac{\alpha}{\alpha_c}}<2,\\
 \Sigma(p^2) &\sim& p^{-1+\sqrt{1-\frac{\alpha}{\alpha_c}}},\\
 \VEV{\bar \psi \psi}_{\Lambda_{ETC}} &\simeq&
     \left( \frac{\Lambda_{ETC}}{\Lambda_{TC}}
     \right)^{1+ \sqrt{1-\frac{\alpha}{\alpha_c}}}
     \cdot \VEV{\bar \psi \psi}_{\Lambda_{TC}}.
\end{eqnarray}
This {\em in principle} can yield more enhancement of the quark mass,
say, up to $O(\Lambda_{TC})$ (if one takes $\gamma_m \simeq 2$).

\begin{flushleft}
  {\bf Top Quark Condensate}\cite{kn:MTY89,kn:Yama92}
\end{flushleft}

Once we have taken a TC with such an extremely tight bound composite
Higgs
with $\gamma_m \simeq 2$, we may consider a much simpler alternative.
As such a top quark condensate (top mode standard model) was proposed
by
Miransky, Tanabashi and Yamawaki (MTY)\cite{kn:MTY89},
based on the \SxSB\ {\em solution of the SD equation} for the
gauged NJL model (this time QCD plus four-fermion interactions)
with $\alpha_{QCD} \ll 1$.  Actually, we find
\begin{eqnarray}
 \gamma_m &\simeq& 1+\sqrt{1-\frac{\alpha_{QCD}}{\alpha_c}}
          \simeq 2- \frac{\alpha_{QCD}}{2\alpha_c}
          \simeq 2- \frac{A}{2 \bar t},\\
 \Sigma(p^2) &\sim& \left(\ln \frac{p}{\Lambda_{QCD}}
 \right)^{-\frac{A}{2}},\\
 \VEV{\bar t t}_{\Lambda} &\simeq&
     \left( \frac{\Lambda}{m_t}\ \right)^2
     \left( \frac{\ln\frac{\Lambda}{\Lambda_{QCD}}}
                 {\ln\frac{m_t}{\Lambda_{QCD}}}
                  \right)^{-\frac{A}{2}}
     \cdot \VEV{\bar t t}_{m_t},
\end{eqnarray}
where we have taken $\mu = m_t$. Using such a very slowly damping
solution
of the SD equation, MTY predicted a rather large top mass
$m_t \simeq 250 {\rm GeV}$ (for $\Lambda \simeq 10^{19} {\rm GeV})$ and
a Higgs boson mass $m_H \simeq 2 m_t$. The idea of the Higgs boson being
a $\bar t t$ composite was also proposed
independently by Nambu\cite{kn:MTY89} in a different context.
An elegant reformulation of the MTY model was further
made by Bardeen, Hill and Lindner (BHL)\cite{kn:MTY89} who newly
included loop
effects of the composite Higgs (and $SU(2) \times U(1)$
gauge interactions) and thereby modified the above MTY prediction into
somewhat smaller value $m_t \simeq 220 {\rm GeV}$. (If we switch off
such
newly included effects in the BHL formulation, we can actually recover
the
original MTY result.)\cite{kn:Yama92}   Although current LEP data seem
to
indicate somewhat smaller value for $m_t$, there have been suggested
many possible variations and modifications reducing the above predicted
value.\cite{kn:Yama92}

\section{Renormalization of the Gauged NJL Model}

All the previous analysis of the gauged NJL model was based on the SD
equation which has an explicit ultraviolet cutoff $\Lambda$. The
continuum limit $\Lambda \rightarrow \infty$ can be taken {\`a la}
Miransky\cite{kn:Mira85} by fine-tuning the {\em bare} coupling(s) to
the critical point (line) so as to keep the fermion dynamical mass
finite.
This defines the $\beta$ function and the anomalous dimension
with respect to the {\em bare} couplings {\em only in the \SxSB phase}
where
the dynamical mass exists. On the other hand, the OPE and the RGE
analysis
we have been talking about should be formulated in terms of the
{\em renormalized} couplings in the continuum theory in the
{\em symmetric phase} as well as the \SxSB phase.
 Actually, it was not until recently
that such a procedure was accomplished by Kondo, Tanabashi and
Yamawaki\cite{kn:KTY91}.  Of course, this renormalization breaks down at
$\alpha \rightarrow 0$ (pure NJL limit) which is obviously
non-renormalizable.

Let us consider QED plus chiral-invariant
 four-fermion interaction
$
  (G_0/2)
$
$
   [(\bar\psi\psi)^2 + (\bar\psi i\gamma_5 \psi)^2]
$
in the ladder approximation as the simplest gauged-NJL model with
a standing gauge coupling, $A \rightarrow \infty$.
By using auxiliary fields $\sigma$ and $\pi$,
we can rewrite the original Lagrangian into an equivalent one;
\begin{equation}
  \Lag   =\bar\psi i\fsl D \psi
        - \bar \psi(\sigma + i\gamma_5 \pi) \psi
  - \frac{1}{G_0} \left[ \frac{\sigma^2+ \pi^2}{2} - m_0 \sigma \right]
        - \frac{1}{4} F_{\mu\nu}F^{\mu\nu},
\label{eq:LagAux}
\end{equation}
with $m_0$ being the bare mass of the fermion.

Now, the multiplicative renormalization can be done phase-independently
through an effective potential written solely in terms of the auxiliary
fields $\sigma$ and $\pi$. Here we use an effective potential obtained
by Bardeen and Love\cite{kn:BL92} (we set $\pi=0$):
\begin{eqnarray}
  -4\pi^2 V(\sigma)
 &=& \Lambda^2 g^{-1} m_0 \sigma
    + \Lambda^2 \left(g^{*}{}^{-1}-g^{-1}\right) \frac{\sigma^2}{2}
\nonumber\\
& & \quad  + \Lambda^4 C \frac{2-\omega}{\alpha/\alpha_c}
    \left( \frac{\sigma}{\Lambda} \right)^{\frac{4}{2-\omega}}
 + \Lambda^4
   {\cal O} ((\frac{\sigma}{\Lambda})^{\frac{4+2\omega}{2-\omega}}
   ,(\frac{\sigma}{\Lambda})^4 ),
\label{eq:BLpot}
\end{eqnarray}
where
\begin{equation}
  g \equiv  \frac{G_0 \Lambda^2}{4\pi^2}
  = {1 \over 4}(1+\omega)^2
  \equiv g^{*},
  \qquad
  \omega \equiv \sqrt{1-\frac{\alpha}{\alpha_c}}
  \qquad (0< \alpha < \alpha_c),
\label{eq:(2.17)}
\end{equation}
and $C$ is a certain constant depending on $\omega$. It is evident that
the phase is determined by the sign of the coefficient of $\sigma^2$
term,
i.e., $g=g^{*}$ is the critical line discoverd by Kondo, Mino and
Yamawaki\cite{kn:KMY89} and by Appelquist, Soldate, Takeuchi and
Wijewardhanai\cite{kn:KMY89} through the ladder SD equation.
It should be emphasized that this potential holds {\em both in
the symmetric and the \SxSB\ phases}.

Now, we renormalize the above effective potential as
follows:\cite{kn:KTY91}
\begin{eqnarray}
  \Lambda^{1-\omega} \sigma & =& \mu^{1 - \omega} \sigma_R,
\label{eq:sigma_r}
\\
\Lambda^{2\omega} (g^{*}{}^{-1}  - g^{-1}  ) &=&
     \mu^{2\omega} (g^{*}_R{}^{-1}- g_R{}^{-1}),
\label{eq:g_r}
\\
 \Lambda^{1+\omega} g^{-1} m_0 &=& \mu^{1+\omega} g_R{}^{-1} m_R,
\label{eq:m_r}
\end{eqnarray}
where $\sigma_R$, $m_R$ (current mass) and $g_R$ are renormalized at the
renormalization point $\mu$. (Note that $g^{*}_R$
($0<g_R^{*}<\infty$) is left arbitrary
and the  simplest choice would be  $g^{*}_R = g^{*}$.)
Through this renormalization we in fact have a renormalized
effective potential $V(\sigma_R)$ ($\Lambda \rightarrow \infty$).
Thus we have obtained a sensible (interacting) continuum theory defined
on the critical line, $g \rightarrow g^*$ as $\Lambda \rightarrow
\infty$.
The theory has been renormalized at zero momentum ($q_\mu=0$) of the
auxiliary field.

As to $q_\mu \ne0$, we need to know the propagator of $\sigma$ which was
calculated by Appelquist, Terning and Wijewardhana\cite{kn:ATW91}.
 Remarkably enough, their $\sigma$  propagator is also renormalized
via the above renormalization condition:\cite{kn:KTY91}
$
 i(4\pi^2)\Delta^{-1}_R (-q^2)
 = -\alpha_c \mu^2 (q^2/\mu^2)^{\omega} /(g^{*} \omega \alpha )
   + \mu^2 (g_R^{*}{}^{-1}-g_R{}^{-1})
$
at $\Lambda \rightarrow \infty$.

 From (\ref{eq:g_r}) and (\ref{eq:m_r}) we immediately
obtain $\beta(g_R)$ and $\gamma_m(g_R)$ by using the $\mu$ independence
of the
bare parameters:\cite{kn:KTY91}
\begin{eqnarray}
  \beta(g_R) &=& 2 \omega g_R \left(1-\frac{g_R}{g^{*}_R} \right),
\label{eq:beta}  \\
  \gamma_m(g_R) &=& 1-\omega + 2\omega \frac{g_R}{g^{*}_R}.
\label{eq:gamma}
\end{eqnarray}

Thus the continuum theory does have a nontrivial UV fixed line
$g_R=g^{*}_R$.
The anomalous dimension at the UV fixed line becomes very large;
\begin{equation}
\gamma_m = 1+ \sqrt{1-\frac{\alpha}{\alpha_c}}
  \qquad (0< \alpha < \alpha_c),
\end{equation}
which thus encompasses a variety of tightly bound composite Higgs
models; walking TC
($\gamma_m \simeq 1$), strong ETC technnicolor ($1<\gamma_m <2$) and
the
top quark condensate ($\gamma_m \simeq 2$).
It is {\em explicitly} checked that the OPE is consistent with this
renormalization and the large anomalous dimension.\cite{kn:KTY91}

\end{document}